\documentclass[a4paper,11pt]{article}
\usepackage{jcappub}
\usepackage[T1]{fontenc}
\title{New cosmographic constraints on the dark energy and dark matter coupling}
\author{Yu.L. Bolotin,}
\author{V.A. Cherkaskiy\note{Corresponding author.}}
\author{and O.A. Lemets}
\affiliation{A.I.Akhiezer Institute for Theoretical Physics, National Science Center ''Kharkov Institute of Physics and Technology'',\\Akademicheskaya Str. 1, 61108 Kharkov, Ukraine}
\emailAdd{ybolotin@gmail.com}
\emailAdd{vcherkaskiy@gmail.com}
\emailAdd{oleg.lemets@gmail.com}
\abstract{We consider three cosmological models with linear interaction between the dark components and obtain restrictions on the coupling constant in terms of the cosmographic parameters. It enables us to find constraints on the coupling constant directly based on observational data and to restrict number of numerous models describing interaction in the dark sector.}
\keywords{cosmographic parameters, interaction in the dark sector}
\arxivnumber{1503.04056}
\begin{document}
\maketitle
\flushbottom
\section{Introduction}
Typically, dark energy (DE) models are based on scalar fields minimally coupled to gravity, and do not implement the explicit coupling of the field to the background matter \cite{bamba,miao}. However there is no fundamental reason for this assumption in absence of underlying symmetry which would suppress the coupling. Given that we do not know the true nature of DE and dark matter (DM), one cannot exclude that there exists a coupling between them. Whereas new forces between DE and baryon matter particles are heavily constrained by observations (e.g. in the Solar system and gravitational experiments on Earth), this is not the case for DM particles. In other words, it is possible that the dark components interact with each other, while not being coupled to standard model particles. The study of the interaction of DE and DM is an important and promising research direction \cite{amendola,zimdahl,bolotin_kostenko_lemets_yerokhin}. Moreover, disregarding the potential existence of an interaction between dark components may result in misinterpretations of observational data \cite{bolotin_erokhin_lemets}. Since there is no fundamental theoretical approach that may specify the functional form of the coupling between DE and DM, presently coupling models are necessarily phenomenological. Of course, one can always provide arguments in favor of a certain type of interaction. However, until the creation of a microscopic theory of dark components, the effectiveness of any phenomenological model will be defined only by how well it corresponds to observations.

Interaction between the dark components is phenomenologically described by the following modification to the conservation equations
\begin{equation}\label{cons_eq}
\dot{\rho }_{dm}+3H\rho _{dm}=Q, \quad \dot{\rho}_{de}+3H(1+w){\rho}_{de}=-Q,
\end{equation}
where $H$ is Hubble parameter, $\rho _{dm}$ and $\rho _{de}$ are densities of DM and DE components respectively. The interaction function $Q$ depends on the scale factor and can take any possible form $Q=Q\left( H,{{\rho }_{dm}},{{\rho }_{de}},t \right).$ However, physically, it makes more sense that the coupling be time-independent with preference given to the factorized-$H$ dependence \(Q=3Hq({{\rho }_{dm}},{{\rho }_{de}}).\)
We consider the simplest linear models with the interaction functions
\begin{equation}\label{if_def}Q_I=3\delta H{{\rho }_{dm}},\quad Q_{II}=3\delta H{{\rho }_{de}},\quad Q_{III}=3\delta H({\rho }_{dm}+\rho_{de}),\end{equation}
which allow exact solutions discussed below, where we find model-independent\footnote{in the sense of independence w.r.t. choice of particular gravity model} constraints on the coupling parameter $\delta$ between DE and DM in the models (\ref{if_def}) formulated exclusively in terms of the cosmographic parameters (CP) values, which can be directly extracted from the cosmological observations.
\section{Cosmography background}
Cosmography is an approach entirely based on the cosmological principle, stating that the Universe is homogeneous and isotropic on scales larger than a hundred megaparsecs. It allows to choose among whole possible variety of models describing the Universe a narrow set of homogeneous and isotropic models. The cosmological principle enables us to build the metrics and make first steps towards interpretation of the cosmological observations. Cosmography is just the kinematics of cosmological expansion. In order to build the key characteristic --- time dependence of the scale factor $a(t)$  --- one needs to take equations of motion (the Einstein's equations) and make an assumption about material content of the Universe, which allow to construct the energy-momentum tensor. Cosmography is efficient because it allows to test any cosmological model which does not contradict the cosmological principle. Modifications of General Relativity or introduction of new components (such as DM and DE) evidently change the dependence $a(t)$ but it does not affect relations between the kinematic characteristics.

In order to make more detailed description of kinematics of cosmological expansion it is useful to consider the extended set of the parameters which includes the Hubble parameter $H(t)\equiv\dot a/a$ and higher order time derivatives of the scale factor \cite{visser1,visser2,capozziello}
$q(t)\equiv -C_2$, $j(t)\equiv C_3$, $s(t)\equiv C_4$, $l(t)\equiv C_5$, where \[ C_n\equiv\frac{1}{a}\frac{{{d}^{n}}a}{d{{t}^{n}}}{{\left[ \frac{1}{a}\frac{da}{dt} \right]}^{-n}}.\]
Dunajski and Gibbons \cite{dunajski_gibbons} proposed an original way to test the cosmological models based on the General Relativity or its modifications. The suggested procedure is the following:
\begin{itemize}
  \item transform the Friedmann equation to ODE for the scale factor; in order to do that, use the conservation equations for each component included in the model to find explicit dependencies of the energy densities on the scale factor;
  \item the obtained ODE includes a row of free parameters (such as initial values of the energy densities and the curvature); let their number equals $N$; differentiate this equation w.r.t. time $N$ times and express higher time derivatives of the scale factor in terms of the cosmographic parameters (CP);
  \item solve the obtained system of $N$ equations w.r.t. the $N$ free parameters and so express them through the CP;
  \item substitute the obtained expressions for all $N$ free parameters into the initial Friedmann equation to obtain the relation between the CP corresponding to the model under consideration; the precision which the obtained relation holds with will determine relevance of the considered model.
\end{itemize}
Such approach can be easily generalized to the models which include such nontrivial effects as interaction between the components, volume and shear viscosity, decay of vacuum energy. Higher CP are presently known with insufficient precision, however perspective of their correction in near future makes the proposed method an efficient tool for testing of cosmological models.
\section{Cosmographic constraints on the coupling constant}
We consider a general case of Universe filled with two components labeled $1$ and $2$ with the EoS and the interaction function respectively
\begin{equation}\label{def}
p_1=w_1\rho_1,\quad p_2=w_2\rho_2,\quad Q=3\delta H{{\rho }_{1}}.
\end{equation}
In particular, the case $Q_I$ in (\ref{if_def}) correspond to $w_1=0$, $w_2=w$ and the case $Q_{II}$ is obtained with $w_1=w$, $w_2=0$ and $\delta\to-\delta$.

Transformation of the independent variable
\[\frac{d}{dt}\equiv aH\frac{d}{da}\] allows to exclude the time variable from (\ref{cons_eq}) and rewrite it in the form
\begin{equation}\label{cons_eqa}
a\rho'_1+3(1+w_1-\delta)\rho_1=0, \quad a\rho'_2+3(1+w_2)\rho_2=-3\delta\rho_1,
\end{equation}
where the prime denotes derivation w.r.t. the scale factor. The system (\ref{cons_eqa}) has exact solution
\[\rho_1(a)=\frac{\rho_{01}}{a^{3(1+w_1-\delta)}}, \quad \rho_2(a)=\frac{\rho_{02}+\delta\rho_{01}/(w_2-w_1+\delta)}{a^{3(1+w_2)}}-\frac{\delta\rho_{01}/(w_2-w_1+\delta)}{a^{3(1+w_1-\delta)}},\]
where $\rho_{0i}\equiv\rho_i(a=1)$, $i=1,2$. Then the first Friedmann equation becomes
\begin{equation}\label{s2}\dot a^2+k=\frac{A}{a^\alpha}+\frac{B}{a^\beta},\end{equation}
where $8\pi G/3\equiv1$, $\alpha\equiv3(w_1-\delta)+1$, $\beta\equiv3w_2+1$, \[A=\rho_{01}\left(1-\frac{1}{1+\frac{\delta}{w_2-w_1}}\right),\quad B=\rho_{02}+\frac{\delta\rho_{01}}{w_2-w_1+\delta}.\]
Then we differentiate the equation (\ref{s2}) two times w.r.t. time to obtain
\begin{align}\label{s3}
\nonumber -2\frac{\ddot a}a\equiv2H^2q &={A}\alpha a^{-\alpha-2} + {B}\beta a^{-\beta-2};\\
2\frac{\dddot a}{\dot a}\equiv2H^2j &={A}\alpha (\alpha+1)a^{-\alpha-2} + {B}\beta (\beta+1)a^{-\beta-2}.
\end{align}
and solve the obtained system of two linear equations w.r.t. $A$ and $B$
\begin{equation}\label{s4}
A =\frac{2H^2(j-(\beta+1)q)}{\alpha(\alpha-\beta)}a^{\alpha+2};\quad
B =\frac{2H^2(j-(\alpha+1)q)}{\beta(\beta-\alpha)}a^{\beta+2}.
\end{equation}
After substitution of (\ref{s4}) into (\ref{s2}) one obtains
\begin{equation}\label{s5}\frac{k}{a^2H^2}=\frac2{\alpha\beta}[q(\alpha+\beta+1)-j]-1.\end{equation}
The third time derivative of the expression (\ref{s2}) leads to
\[\frac2H\left(\frac{a^{IV}}{\dot a}-\frac{\ddot{a}\ \dddot{a}}{\dot a^2}\right)\equiv2H^2(s+qj)= -\frac{A\alpha(\alpha+1)(\alpha+2)}{a^{\alpha+2}} - \frac{B\beta(\beta+1)(\beta+2)}{a^{\beta+2}}.\]
Using the expressions (\ref{s4}), the latter equation can be transformed to the form
\begin{equation}\label{s6}
s+qj+(\alpha+\beta+3)j-q(\alpha+1)(\beta+1)=0.
\end{equation}
Substituting the explicit expressions for $\alpha$ and $\beta$, (\ref{s5}) and (\ref{s6}) take on the form
\begin{equation}
\label{s50}\frac{k}{a^2H^2}=\frac{2[3q(w_2+w_1-\delta+1)-j]}{(3w_2+1)[3(w_1-\delta)+1]}-1\\
\end{equation}
\begin{equation}
\label{s60}s+qj+[3(w_2+w_1-\delta)+5]j-q(3w_2+2)[3(w_1-\delta)+2]=0.
\end{equation}
From (\ref{s60}) one obtains $\delta$ as explicit function of CP and EoS parameters
\begin{equation}
\label{delta1}\delta=2+3w_1+\frac{s+qj+3(w_2+1)j}{j-(3w_2+2)q}.
\end{equation}
Under the assumption of flatness ($k=0$), (\ref{s50}) gives an alternative relation
\begin{equation}
\label{delta0}\bar\delta=\frac13+w_1+\frac{j-q(3w_2+2)}{3[(3w_2+1)/2-q]}.
\end{equation}
For the models $Q_I$ and $Q_{II}$ in (\ref{if_def}) one obtains respectively
\begin{align}
\label{deltai}\bar\delta_I&=\frac13+\frac23\frac{j-q(3w+2)}{3w+1-2q},& \delta_I&=2+\frac{s+qj+3(w+1)j}{j-(3w+2)q}; \\
\label{deltaii}\bar\delta_{II}&=\frac13+w+\frac23\frac{j-2q}{1-2q},& \delta_{II}&=2+3w+\frac{s+(q+3)j}{j-2q}.
\end{align}
\section{Test of the models with the observational data}
\begin{figure}[tbp]
\centering
\includegraphics[width=.9\textwidth]{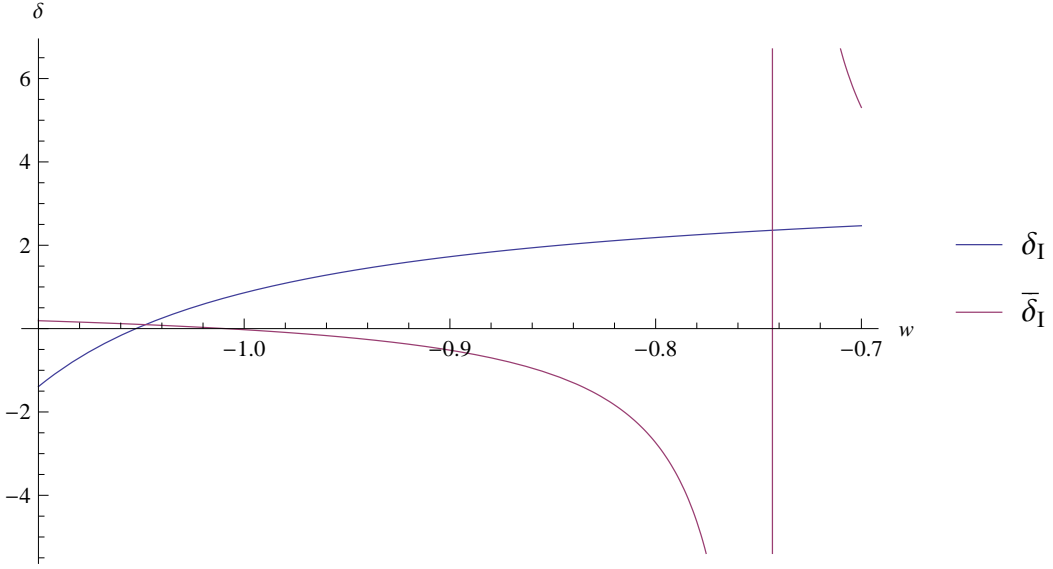}
\caption{\label{q1} The EoS parameter dependence of the coupling constants $\delta$ and $\bar\delta$ in the model $Q_I$ (\ref{deltai}).}
\end{figure}

\begin{figure}[tbp]
\centering
\includegraphics[width=.9\textwidth]{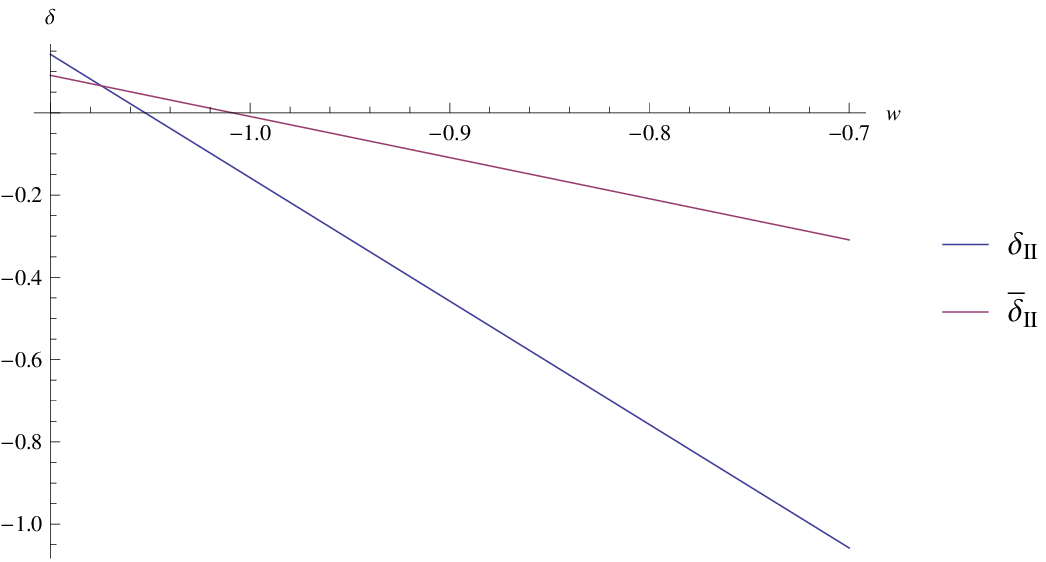}
\caption{\label{q2} The EoS parameter dependence of the coupling constants $\delta$ and $\bar\delta$ in the model $Q_{II}$  (\ref{deltaii}).}
\end{figure}

We plot $\bar\delta_{I,II}$ and $\delta_{I,II}$ (see figures \ref{q1} and \ref{q2}) for $-1.1<w<-0.7$ (it is the present observational estimate on the EoS parameter for DE), taking the CP values $H_0=74.22_{-5.08}^{+5.23}$, $q_0=-0.6149_{-0.2238}^{+0.2716}$, $j_0=1.030_{-1.001}^{+0.722}$, $s_0=0.16_{-1.03}^{+1.45}$ from \cite{1204.2007}, where the authors use the data of the SNeIa Union 2 compilation by the supernovae cosmology project \cite{42}.

\begin{figure}[tbp]
\centering
\includegraphics[width=.9\textwidth]{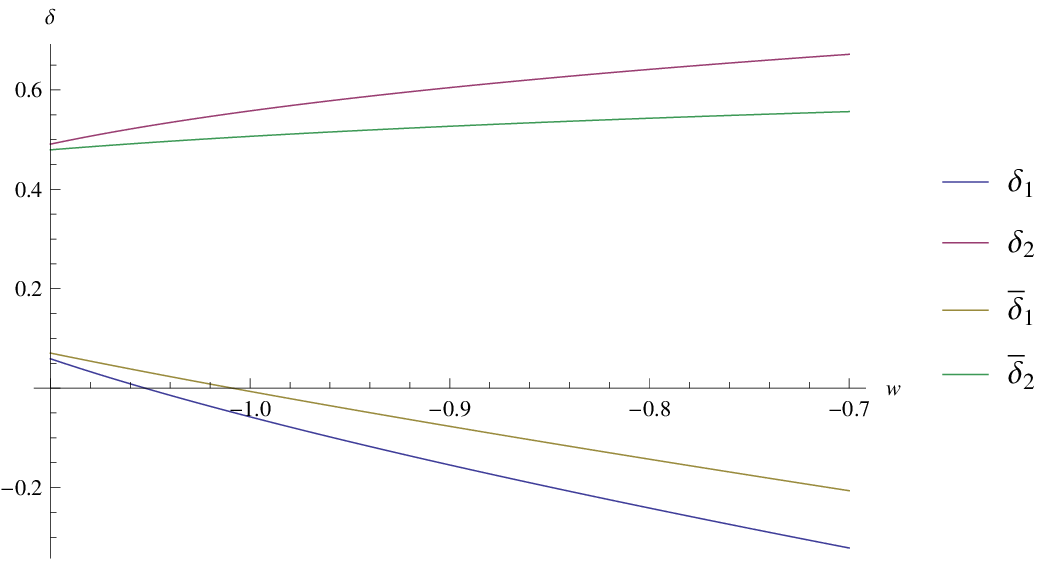}
\caption{\label{q3}The EoS parameter dependence of the coupling constants $\delta$ (\ref{delta3}) and $\bar\delta$ (\ref{delta30}) in the model $Q_{III}$.}
\end{figure}

The model $Q_{III}$ also has exact but cumbersome solution:
\begin{align}
\label{dm3}\rho_{dm}=\frac{\rho_{dm0}}2\left(\frac1{a^{\alpha+2}}+\frac1{a^{\beta+2}}\right)&+ \frac{\rho_{de0}-\left(1+\frac{w}{2\delta}\right)\rho_{dm0}}{2\sqrt{1+\left(1+\frac{w}{2\delta}\right)^2}}\left(\frac1{a^{\alpha+2}}-\frac1{a^{\beta+2}}\right),\\
\label{de3}\rho_{de}=\frac{\rho_{de0}}2\left(\frac1{a^{\alpha+2}}+\frac1{a^{\beta+2}}\right)&- \frac{\rho_{dm0}-\left(1+\frac{w}{2\delta}\right)\rho_{de0}}{2\sqrt{1+\left(1+\frac{w}{2\delta}\right)^2}}\left(\frac1{a^{\alpha+2}}-\frac1{a^{\beta+2}}\right),
\end{align}
where \[\{\alpha,\beta\}\equiv1+3\delta\left[\frac{w}{2\delta}\pm\sqrt{1+\left(1+\frac{w}{2\delta}\right)^2}\right].\]
The corresponding Friedmann equation again has form (\ref{s2}), the relations (\ref{s5}) and (\ref{s6}) hold and give quadratic equations for $\delta$, so one finally obtains (see figure \ref{q3})
\begin{align}
\label{delta3}\delta_{1,2}&=-\frac w 4\left\{1\mp\sqrt{1-\frac8{9qw^2}\left[s - 2 q (2 + 3 w) + j (5 + q + 3 w)\right]}\right\}\\
\label{delta30}\bar\delta_{1,2}&=-\frac w 4\left\{1\mp\sqrt{1-\frac8{9w^2}\left[1 + 2 j + 3 w - 6 q (1 + w)\right]}\right\}.
\end{align}
Careful analysis of the plots reveals a remarkable feature: the functions $\delta(w)$ corresponding to different models ($I,II,III$) have exact common roots, namely:
\begin{align}
\delta_I(w_{01})=\delta_{II}(w_{01})=\delta_1(w_{01})=0,\quad & w_{01}=-\frac{s+(q+5) j - 4 q }{3 (j - 2 q)}\approx-1.06;\\
\bar\delta_I(w_{02})=\bar\delta_{II}(w_{02})=\bar\delta_1(w_{02})=0,\quad & w_{02}=-\frac{1+2 j -6 q }{3 (1 - 2 q)}\approx-1.009.
\end{align}
For the case of DE in form of cosmological constant ($w=-1$) the corresponding estimates for the coupling constant $\delta$ are the following:
\begin{align}
\delta_I(-1)&\approx0.86, & \delta_{II}(-1)&\approx-0.16, & \delta_1(-1)&\approx-0.06; \\
\bar\delta_I(-1)&\approx-0.026, &\bar\delta_{II}(-1)&\approx-0.00897, &\bar\delta_1(-1)&\approx-0.0066.
\end{align}
We disregard the solutions $\delta_2$ and $\bar\delta_2$ from (\ref{delta3}) and (\ref{delta30}) respectively as they give unreasonable results for the coupling constant.

\section{Summary}
We considered three cosmological models with linear type of interaction between the dark components and obtained two types of restrictions on the coupling constant $\delta$ in terms of the CP: the first type contains the CP $q$, $j$ and $s$, the second is obtained in assumption that the Universe is flat ($k=0$) and contains only $q$ and $j$. We resolved the obtained constraints to explicitly obtain $\delta$ as a function of CP and EoS parameter $w$ and plotted the dependence in the interval $-1.1<w<-0.7$ which corresponds to the present observations. We see that all the three considered models give similar results for the coupling constant $\delta$, with exceptions of the model $Q_I$ where $\bar\delta(w)$ diverges in the considered interval (for $w\approx-0.74$). Numerical estimates of the coupling constant for the case of cosmological constant ($w=-1$) for all the three considered models give small negative value of $\delta$, with exceptions of the model $Q_I$ where $\delta\approx1$.

\acknowledgments{We are grateful to D. A. Yerokhin for useful discussions.}

\end{document}